\documentclass[preprint,onecolumn,nofootinbib]{revtex4}
\usepackage[colorlinks=true,linkcolor=blue,urlcolor=blue,filecolor=black,citecolor=red,pdfstartview=FitV,pdftitle={},pdfsubject={},pdfkeywords={},pdfpagemode=None,bookmarksopen=true]{hyperref}
\usepackage{graphicx}
\usepackage{amsmath}
\usepackage{amsfonts}
\usepackage{amssymb,ulem}
\usepackage{color,xcolor}%
\usepackage{CJK}
\usepackage{subfigure}
\usepackage{amsthm,amsmath,amssymb}
\usepackage{mathrsfs}
\usepackage{multirow}

\usepackage{dcolumn}
\usepackage{float}
\newcommand{\delay}[1]{\mathbf{D}_{#1}}

\begin{document}
\title{Probing Quantum Gravity effects with Extreme Mass Ratio Inspirals around Rotating Hayward Black Holes}
\author{Dan Zhang $^{1}$}
\thanks{danzhanglnk@163.com}
\author{Chao Zhang $^2$}
\thanks{zhangchao1@nbu.edu.cn, corresponding author}
\author{Qiyuan Pan$^{3}$}
\thanks{panqiyuan@hunnu.edu.cn}
\author{Guoyang Fu$^{1}$}
\thanks{fuguoyang@yzu.edu.cn}
\author{Jian-Pin Wu$^{1}$}
\thanks{jianpinwu@yzu.edu.cn, corresponding author}
\affiliation{$^1$\mbox{Center for Gravitation and Cosmology, College of Physical Science and Technology,} \mbox{Yangzhou University, Yangzhou 225009, China}\\
$^2$Institute of Fundamental Physics and Quantum Technology, Department of Physics, School of Physical Science and Technology, Ningbo University, Ningbo, Zhejiang 315211, China \\
$^3$Department of Physics, Key Laboratory of Low Dimensional Quantum Structures and Quantum Control of Ministry of Education, \mbox{ Synergetic Innovation Center for Quantum Effects and Applications,} \mbox{ Hunan Normal University, Changsha, 410081, Hunan, China}}

\begin{abstract}

We investigate extreme mass-ratio inspirals (EMRIs) around a rotating Hayward black hole to assess the detectability of signatures arising from quantum gravity.
The quantum parameter $\alpha_0$, which encodes deviations from general relativity (GR), introduces extra correction terms in both the orbital frequency and the fluxes.
Our results show that after one year of accumulated observation, these corrections induce a detectable dephasing in the EMRI waveform. 
Using the modified orbital evolution driven by $\alpha_0$, we generate waveforms via the augmented analytic kludge (AAK) model implemented in the \texttt{FastEMRIWaveforms} package. 
Furthermore, we utilize the time-delay interferometry (TDI) to suppress the laser noise and phase fluctuations induced by spacecraft motion, and then employ the Fisher information matrix (FIM) to test the sensitivity of LISA in detecting deviations from GR. Our results demonstrate the potential of LISA to probe quantum-gravity effects through high-precision observations of EMRIs.
\end{abstract}

\maketitle

\section{Introduction}  \label{sec-intro}

The successful detection of gravitational waves (GWs) marked the dawn of GW astronomy. To date, ground-based detectors have observed hundreds of high-frequency GW events ($10 \sim10^3$ Hz) generated by mergers of compact objects such as binary black holes (BH) and binary neutron stars, opening a new window to understand fundamental physics \cite{LIGOScientific:2016emj,LIGOScientific:2017vwq,LIGOScientific:2018mvr,LIGOScientific:2020ibl,LIGOScientific:2021usb,KAGRA:2021vkt}.
In the future, space-based GW observatories, like LISA \cite{LISA:2017pwj,LISA:2024hlh}, TianQin \cite{Luo:2015ght,TianQin:2020hid,Gong:2021gvw}, and Taiji \cite{Hu:2017mde,Gong:2021gvw}, will access the millihertz band, thereby bridging a crucial gap in the GW spectrum.

Extreme mass ratio inspirals (EMRIs) are among the most promising sources for space-based GW observatories, with an expected detection rate of several to thousands per year \cite{Gair:2004iv,Mapelli:2012pw,Babak:2017tow,Bonetti:2020jku,Pozzoli:2023kxy,Fan:2024nnp}.
Characterized by small extreme mass ratio ($ \lesssim 10^{-4}$), an EMRI signal can complete up to $\sim 10^5$ orbital cycles within the LISA sensitivity band \cite{Amaro-Seoane:2007osp,Barack:2006pq,Canizares:2012is,Moore:2017lxy,Das:2025eiv}. 
While this long signal duration poses significant challenges for waveform modeling and data analysis \cite{Hughes:2021exa,Wardell:2021fyy,Iglesias:2025tnt,Zhang:2025eqz,Wei:2025lva,Gair:2008zc,MockLISADataChallengeTaskForce:2009wir,Baghi:2022ucj,Li:2024rnk,Fan:2020zhy,Speri:2025ucn,Xiaobo:2025jkw}, EMRIs nevertheless provide an exquisitely sensitive laboratory for testing theories of gravity in the strong-field regime. The rich information encoded in EMRI waveforms makes them powerful probes for fundamental physics, which can be used to search for new fundamental fields \cite{Maselli:2021men,Barsanti:2022vvl,Barsanti:2022ana,Collodel:2021jwi,DellaRocca:2024pnm,Fell:2023mtf,Zhang:2024ogc,Zhang:2022rfr,Guo:2022euk,Zhang:2022hbt,Liang:2022gdk,Zi:2025jxy,Zi:2025qos,Zi:2024lmt}, constrain modified gravity theories \cite{Zhang:2024csc,Qiao:2024gfb,AbhishekChowdhuri:2023gvu,Lu:2025xlp,Xia:2025yzg,Yunes:2011aa,Guo:2023mhq,Sopuerta:2009iy,Pani:2011xj}, and even probe potential quantum gravity effects \cite{Fu:2024cfk,Zi:2024jla,Liu:2024qci,Yang:2025esa,Yang:2024cnd,Yang:2024lmj,Ahmed:2025shr,Gong:2025mne}.
Furthermore, there is also growing interest in using EMRIs to study the astrophysical environment of supermassive black holes (SMBH), including dark matter halos \cite{Duque:2023seg,Cardoso:2022whc,Figueiredo:2023gas,Zhang:2024ugv,Gliorio:2025cbh,Rahman:2023sof,Dai:2023cft,Zhao:2024bpp,Mitra:2025tag} and accretion disks \cite{Destounis:2022obl,Barausse:2007dy,Yunes:2011ws,Kocsis:2011dr,Duque:2024mfw}.
Collectively, these theoretical works establish a vital theoretical framework for the future GW astronomy, highlighting EMRIs as a cornerstone target for upcoming space missions.

LISA data analysis confronts several distinct challenges. 
The first is non-stationary detector noise, known as “glitches”, primarily caused by spacecraft outgassing \cite{LISAPathfinder:2022awx,Sala:2023hpr}, which can introduce measurable biases in EMRI parameter estimation \cite{Boumerdassi:2025gvf,Sauter:2025iey,Spadaro:2023muy,Castelli:2024sdb}.
The second is the overlap of signals from multiple GW sources, a problem commonly tackled through techniques like global fitting \cite{Littenberg:2023xpl,Katz:2024oqg,Deng:2025wgk}.
A further challenge stems from the spacecraft's inability to maintain perfectly equal arm lengths due to perturbations from planetary gravity and orbital correction thrusts \cite{Frank:2020tjx}.
To suppress the resulting laser frequency noise, the response function must account for dependencies on both frequency and time.
This is achieved by constructing Time-Delay Interferometry (TDI) observables using precise time shifts and delays \cite{Cornish:2002rt,Cornish:2003tz,Tinto:1999yr,Tinto:2002de,Tinto:2014lxa,Vallisneri:2004bn,Marsat:2018oam,Armstrong:1999er, Estabrook:2000ef}.
The first-generation TDI has been widely adopted in space-based GW data analysis \cite{Tinto:2001ii,Tinto:2001ui,Prince:2002hp,Armstrong:2001uh,Tinto:2003uk,Tinto:2004nz,Zhang:2020khm}, with second-generation methods under active discussion \cite{Krolak:2004xp,Vallisneri:2005ji,Wang:2017aqq,Wang:2020fwa,RajeshNayak:2004jzp,Nayak:2005un,Zhang:2025fiu}.

The Hawking–Penrose singularity theorem demonstrates that gravitational collapse within the framework of general relativity (GR) inevitably leads to the formation of spacetime singularities, revealing the incompleteness of the theory in high-curvature regions of spacetime \cite{Penrose:1964wq}.
Quantum gravity is expected to resolve this fundamental issue. At the phenomenological level, one approach is to introduce exotic matter fields that replace the central singularity with a nonsingular core of finite curvature.
The asymptotic behavior of such a core typically classifies these regular BH models into two main types: those with a de Sitter core \cite{Ayon-Beato:1998hmi,2767662,Hayward:2005gi,Boos:2023icv} and those with a Minkowski core \cite{Xiang:2013sza,Culetu:2014lca,Simpson:2019mud}. As highlighted in Refs. \cite{Hayward:2005gi, Pedraza:2020uuy}, those regular black hole models play a significant role in understanding black hole formation and evaporation. With the advent of groundbreaking GW astronomy, probing the quantum gravity imprints from the GW signals has emerged as a major research frontier. Theoretically, a series of studies \cite{Konoplya:2022pbc,Fu:2023drp,Bolokhov:2023ozp,Konoplya:2023ahd,Gong:2023ghh,Zhu:2024wic,Konoplya:2024lch,Zhang:2024nny,Zhang:2025ygb,Dong:2024ams,Shi:2025gst,Tang:2024txx,Konoplya:2025hgp} have investigated the properties of quasinormal modes (QNMs) of the effective quantum gravity model, showing that a slight near-horizon deformation induced by quantum-gravity effects can trigger a pronounced outburst in the overtones. Such overtone outbursts are expected to become observable by future space-based gravitational-wave detectors \cite{LIGOScientific:2025rid,LIGOScientific:2025obp}.

In this work, we aim to investigate the potential of EMRIs to probe quantum gravity effects. To this end, we adopt a rotating Hayward regular black hole (HBH) as the background spacetime for the central massive object. This choice is motivated by several interrelated considerations. First, among various regular BH models, the Hayward metric, which directly replaces the central singularity with a Planck-scale non-singular core, is particularly attractive due to its mathematical simplicity. It is characterized by a single parameter $\alpha_0$ that quantifies deviations from GR, greatly facilitating the parametrization and integration of such corrections into waveform models. Second, astrophysical supermassive BHs are expected to possess significant angular momentum. Rotation profoundly alters spacetime geometry, influencing both orbital dynamics and GW emission. Studying the rotating version of a regular BH is essential for astrophysical realism. In particular, as shown later in Fig.~\ref{phase_diagram}, the rotating HBH derived via the Newman-Janis algorithm preserves the non-singular core and finite curvature invariants, and more importantly introduces a coupling between the quantum parameter $\alpha_0$ and the rotation parameter $a$, a key feature absent in static regular HBHs. Notably, rotational effects are expected to amplify the influence of quantum corrections on orbital frequencies and GW fluxes, thereby enhancing the detectability of such subtle deviations with future observatories. Finally, EMRI systems, with their numerous orbital cycles and long-duration observational windows (months to years), provide a unique ``natural laboratory'' in which tiny deviations in the background spacetime can accumulate into measurable phase shifts in the gravitational waveform. In contrast to short-lived QNMs, rotating HBH’s quantum deformations, combined with the long inspiral timescale of EMRIs, produce cumulative dephasing detectable by LISA. In summary, studying EMRIs around a rotating HBH offers a theoretically consistent and astrophysically realistic framework for systematically evaluating the detectability of quantum gravity effects through GW observations.

This paper is organized as follows. In Section \ref{sec-1}, we provide a brief introduction to the rotating HBH, with a particular focus on analyzing the dependence of its phase structure on the parameters $a$ and $\alpha_0$. Section \ref{sec-2} presents the expressions for the quantum-corrected geodesic orbits and the fluxes. Subsequently, in Section \ref{sec-3}, we compute the orbital evolution and generate the EMRI waveforms using the \texttt{FastEMRIWaveforms} package. The Time-Delay Interferometry (TDI) technique is employed to suppress laser noise, and the Fisher Information Matrix (FIM) is used to assess the LISA detector's sensitivity to quantum gravity effects in Section \ref{sec-4}. Finally, we summarize our findings and discuss future prospects in Section \ref{sec-5}.

\section{Rotating Hayward black hole}\label{sec-1}

In this section, we begin with the HBH, which was first proposed by Hayward in \cite{Hayward:2005gi}. The metric for a static, spherically symmetric HBH is given by
\begin{eqnarray} \label{metric_sph}
ds^2=-f(r)dt^2+f(r)^{-1}dr^2 +r^2(d\theta^2+\sin^2\theta d\phi^2)\,,
\end{eqnarray}
where the shape function $f(r)$ reads
\begin{eqnarray} 
f(r)=1-\frac{2m(r)}{r}\,,  \ \ m(r)=\frac{M r^3}{r^3+\alpha_0 M}.
\end{eqnarray}
Here, $\alpha_0$ is a regularization parameter characterizing the deviation from the Schwarzschild black hole. Consequently, the metric \eqref{metric_sph} reduces to the Schwarzschild case when $\alpha_0 = 0$.
The rotating generalization of this black hole was first derived by Bambi and Modesto via the Newman-Janis algorithm \cite{Newman:1965tw,Newman:1965my,Drake:1998gf}, which reads \cite{Bambi:2013ufa}
\begin{eqnarray} \label{metric}
ds^2&=&-\left(1-\frac{2m(r)r}{\Sigma}\right)dt^2-\frac{4a m(r)r \sin^2{\theta}}{\Sigma}dtd\phi+\frac{\Sigma}{\Delta}dr^2 \nonumber \\
&&+\Sigma d\theta^2+\left(r^2+a^2+\frac{2a^2m(r)r\sin^2{\theta}}{\Sigma}\right)\sin^2{\theta}d\phi^2\,,
\end{eqnarray}
with
\begin{eqnarray} 
\Delta=r^2-2m(r)r+a^2,  \ \ \Sigma=r^2+a^2\cos^2\theta,
\end{eqnarray}
where $a$ is the rotation parameter. This new spacetime preserves the curvature invariants of the non-rotating solution at the origin. The horizons are located at the roots of $\Delta=0$, which imposes a constraint on the permissible range of the parameters $\{a/M, \alpha_0/M^2\}$ to avoid the horizonless. 

\begin{figure}[H]
	\centering
	\includegraphics[width=0.45\textwidth]{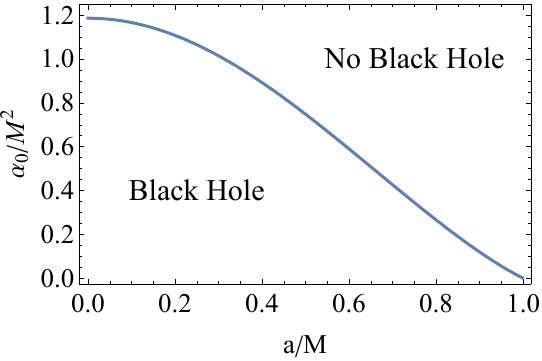}\hspace{0.2mm}	
	\caption{The phase diagram of the rotating HBH over $\{a/M, \alpha_0/M^2\}$. The blue line represents the case of extremal black hole. }
	\label{phase_diagram}
\end{figure}

Fig. \ref{phase_diagram} displays the phase structure of the rotating HBH in the $(a, \alpha_0)$ parameter space. As the spin parameter $a$ increases, the allowable value of the parameter $\alpha_0$ decreases as $a$ increases. In the non-rotating limit ($a=0$), this bound reduces to $\alpha_0 < 32/27$, consistent with the static case. Notably, while preserving the non-singular core and finite curvature invariants of its static counterpart, the rotating generalization introduces a significant new feature: a non-trivial coupling between the quantum parameter $\alpha_0$ and the spin parameter $a$, a key characteristic absent in the static regular HBH model.

\section{Geodesic and Fluxes}\label{sec-2}

We restrict the motion of the test particle to equatorial orbits by setting $\theta = \pi/2$, which forces the Carter constant to vanish, i.e., $\mathcal{Q}=0$ \cite{Hughes:1999bq, Glampedakis:2002ya, Glampedakis:2002cb, AbhishekChowdhuri:2023gvu}. The associated conserved quantity of the energy $E$ and angular momentum $L_z$ can be expressed as 
\begin{eqnarray} 
\label{En}
E&=&-u_\mu \xi^\mu{(t)}=-g_{tt}u^t-g_{t\phi}u^\phi , \\
\label{Lz}
L_z&=&u_\mu \xi^\mu{(\phi)}=g_{t\phi}u^t+g_{\phi\phi}u^\phi,
\end{eqnarray}
where the Killing vectors $\xi^\mu_{(t)}=(1,0,0,0)$ and $\xi^\mu_{(\phi)}=(0,0,0,1)$, and $u^\mu$ is the four-velocity for the test particle. For convenience, we introduce the following dimensionless quantities: $\hat{r}=r/M, \hat{a}=a/M, \hat{\alpha}_0=\alpha_0/M^2, \hat{L}_z=L_z/(\mu M) \ \text{and} \ \hat{E}=E/\mu $. Henceforth, we set $M=1$ and drop the hats on all dimensionless variables.

The dynamics are governed by the conserved energy $E$ and angular momentum $L_z$ through the geodesic equations
\begin{eqnarray} \label{Geodesic}
&&r^2\frac{dt}{d\tau}=\frac{(r^2+a^2)P}{\Delta}-a(a E-L_z), \nonumber \\
&&r^4\left(\frac{dr}{d\tau}\right)^2=P^2-\Delta(r^2+(a E-L_z)^2), \\
&&r^2\frac{d\phi}{d\tau}=\frac{a P}{\Delta}-(a E-L_z), \nonumber 
\end{eqnarray}
where $P=E(r^2+a^2)-a L_z$.
To describe eccentric motion, we parametrize bound orbits using the eccentricity $e$ and semi-latus rectum $p$ as
\begin{eqnarray} 
r=\frac{p}{1+e\cos\chi}.
\end{eqnarray}
Therefore, $E$ and $L_z$ can be solved as
\begin{eqnarray}\label{MoC}
\begin{split}
E&=\frac{\sqrt{2e}\sqrt{p+2(e-1)m(r_a)}\sqrt{p-2(e+1)m(r_p)}}{\sqrt{p}\sqrt{2ep-(e-1)^3m(r_a)-(e+1)^3m(r_p)}}   \\
&-a \frac{(e^2-1)^2\left[(e-1)m(r_a)+(e+1)m(r_p)\right]^{3/2}}{p\left[2ep-(e-1)^3m(r_a)-(e+1)^3m(r_p)\right]^{3/2}}, \\
L_z&=\frac{p\sqrt{(e-1)m(r_a)+(e+1)m(r_p)}}{\sqrt{2ep-(e-1)^3m(r_a)-(e+1)^3m(r_p)}}  \\
&- a \frac{\sqrt{2e}\sqrt{p+2(e-1)m(r_a)}\sqrt{p-2(e+1)m(r_p)}((e-1)^3m(r_a)+(e+1)^3m(r_p))}{\sqrt{p}\left[2ep-(e-1)^3m(r_a)-(e+1)^3m(r_p)\right]^{3/2}},
\end{split}
\end{eqnarray}
where $r_a=p/(1-e)$ and $r_p=p/(1+e)$. When the $m(r_{a})=m(r_p)=1$, the expression of \eqref{MoC} reverts to the Kerr BH case.

For the eccentric equatorial orbit, the radial and azimuthal motion are characterized by two fundamental frequencies $(\Omega_r,\Omega_\phi)$, given by
\begin{eqnarray} 
\label{frequency}
\Omega_\phi&=&\frac{L_z r+2(a E -L_z) m(r)}{E r^3-2a L_z m(r)}, \nonumber\\
\Omega_r&=&\frac{2\pi}{T_r}; \ \  T_r=\int_{0}^{2\pi}d\chi \frac{dt}{d\chi},
\end{eqnarray}
where $T_r$ denotes the radial time period. Recalling  Eqs. \eqref{Geodesic} and \eqref{MoC}, the frequencies defined in \eqref{frequency} are given by
\begin{eqnarray} 
\label{frequency_exp}
\Omega_\phi&=&\Omega_\phi^{GR}+\alpha_0 \Omega_\phi^{QC}, \nonumber\\
\Omega_r&=&\Omega_r^{GR}+\alpha_0 \Omega_r^{QC}.
\end{eqnarray}
Here $\Omega_\phi^{GR}$ and $\Omega_r^{GR}$ denote the frequencies of the GR, which can be obtained from the \texttt{KerrGeodesics} package \cite{KerrGeodesics, KerrAngleConversions} in the Black Hole Perturbation Toolkit \cite{BHPToolkit}. The terms $\Omega_\phi^{QC}$ and $\Omega_r^{QC}$ then represent the quantum-corrected contributions, defined as the difference after subtracting the GR contribution. Their expressions are: 
\begin{eqnarray} 
\label{frequency_delta}
\Omega_\phi^{QC}&=&-\frac{(6-2\sqrt{1-e^2}+e^2(9+2\sqrt{1-e^2}))(1-e^2)^{3/2}}{2p^{9/2}}\nonumber\\ 
&-&\frac{a(90+62\sqrt{1-e^2}+3e^2(17+4\sqrt{1-e^2}))(1-e^2)^{5/2}}{2p^6}, \nonumber\\
\Omega_r^{QC}&=&\frac{(3+\sqrt{1-e^2})(1-e^2)^{5/2}}{p^{9/2}}\\
&-&\frac{3a(5+9\sqrt{1-e^2}+e^2(7+2\sqrt{1-e^2}))(1-e^2)^{5/2}}{p^6}. \nonumber
\end{eqnarray}

The GW radiation carries away the orbital energy and angular momentum. This loss can be decomposed into distinct contributions from GR and quantum corrections
\begin{eqnarray} 
\label{fluxes_total}
\left<\frac{dE}{dt}\right>&=&\left<\frac{dE}{dt}\right>^{GR}+\alpha_0 \left<\frac{dE}{dt}\right>^{QC}, \nonumber\\
\left<\frac{dL_z}{dt}\right>&=&\left<\frac{dL_z}{dt}\right>^{GR}+\alpha_0 \left<\frac{dL_z}{dt}\right>^{QC},
\end{eqnarray}
where $\left\langle {dE}/{dt} \right\rangle^{GR}$ and $\left\langle {dL_z}/{dt} \right\rangle^{GR}$ build upon higher-order post-Newtonian results, augmented by phenomenological corrections derived from Teukolsky-based fits to the angular momentum and inclination evolution \cite{Gair:2005ih,Glampedakis:2002cb}. For the quantum-corrected contributions, we adopt the quadrupole-octupole formula \cite{Babak:2006uv} to calculate the corresponding fluxes $\left\langle {dE}/{dt} \right\rangle^{QC}$ and $\left\langle {dL_z}/{dt} \right\rangle^{QC}$, which are expressed as
\begin{eqnarray} 
\label{quad-octu}
\left<\frac{dE}{dt}\right>^{QC}&=&-\frac{1}{5}\left<I_{STF}^{ij(3)}I_{STF}^{ij(3)}+\frac{5}{189}M_{STF}^{ijk(3)}M_{STF}^{ijk(3)}+\frac{16}{9}J_{STF}^{ij(3)}J_{STF}^{ij(3)}\right> \nonumber ,\\
\left<\frac{dL_z}{dt}\right>^{QC}&=&-\frac{2}{5}\epsilon^{zkl}\left<I_{STF}^{kj(3)}I_{STF}^{jl(3)}+\frac{5}{126}M_{STF}^{kjn(3)}M_{STF}^{ljn(3)}+\frac{16}{9}J_{STF}^{kj(3)}J_{STF}^{lj(3)}\right>,
\end{eqnarray}
where the number of the superscript denote the derivatives with the time $t$, and the mass quadrupole moment $I^{ij}$, current quadrupole moment $J^{ij}$ and mass octupole moment $M^{ijk}$ are given by \cite{Babak:2006uv}:
\begin{eqnarray} 
\label{quadrupole-octupole}
I^{ij}&=&\mu x_p^i x_p^j, \nonumber \\
J^{ij}&=&\epsilon^{jlm} v^m I^{li}, \\
M^{ijk}&=&x^i I^{jk}. \nonumber
\end{eqnarray}
In the weak-field approximation, the QC fluxes of the energy and angular momentum are obtained as 
\begin{eqnarray} 
\label{fluxes}
\left<\frac{dE}{dt}\right>^{QC}&=&-(1-e^2)^{3/2}(E_1+e^2 E_2+e^4 E_3+e^6 E_4)p^{-8} \nonumber \\ 
&&+a(1-e^2)^{3/2}(E_5+e^2 E_6+e^4 E_7+e^6 E_8+e^8 E_9)p^{-9.5}, \\
\left<\frac{dL_z}{dt}\right>^{QC}&=&-(1-e^2)^{3/2}(L_1+e^2 L_2+e^4 L_3+e^6 L_4)p^{-6.5} \nonumber \\
&&+a(1-e^2)^{3/2}(L_5+e^2 L_6+e^4 L_7+e^6 L_8+e^8 L_9)p^{-8}, 
\end{eqnarray}
with
\begin{eqnarray} 
E_1&=&\frac{416}{5}-\frac{32}{5}\sqrt{1-e^2}, \ E_2=\frac{10804}{15}-\frac{196}{15}\sqrt{1-e^2}, \ E_3=\frac{2524}{3}+17\sqrt{1-e^2}, \nonumber \\
E_4&=&\frac{3779}{30}+\frac{37}{15}\sqrt{1-e^2}, \ E_5=776-\frac{1464}{15}\sqrt{1-e^2}, \ E_6=\frac{132896}{15}-\frac{8876}{15}\sqrt{1-e^2}, \nonumber \\
E_7&=&\frac{267728}{15}+\frac{7507}{15}\sqrt{1-e^2}, \ E_8=\frac{48176}{5}+\frac{8953}{30}\sqrt{1-e^2}, \ E_9=\frac{48031}{48}+\frac{713}{30}\sqrt{1-e^2}, \nonumber \\
L_1&=&\frac{352}{5}-\frac{32}{5}\sqrt{1-e^2}, \ L_2=\frac{1636}{5}+\frac{4}{5}\sqrt{1-e^2}, \ L_3=\frac{757}{5}+\frac{28}{5}\sqrt{1-e^2},  \\
L_4&=&\frac{12}{5}, \ L_5=\frac{9224}{15}-\frac{3368}{15}\sqrt{1-e^2}, \ L_6=\frac{63104}{15}-\frac{868}{15}\sqrt{1-e^2}, \nonumber \\
L_7&=&\frac{23864}{5}+229\sqrt{1-e^2}, \ L_8=\frac{12087}{10}+\frac{267}{5}\sqrt{1-e^2}, \ L_9=\frac{84}{5}.\nonumber
\end{eqnarray}

\section{orbital evolution and waveform}\label{sec-3}

\begin{figure}[H]
	\centering
	\includegraphics[width=0.45\textwidth]{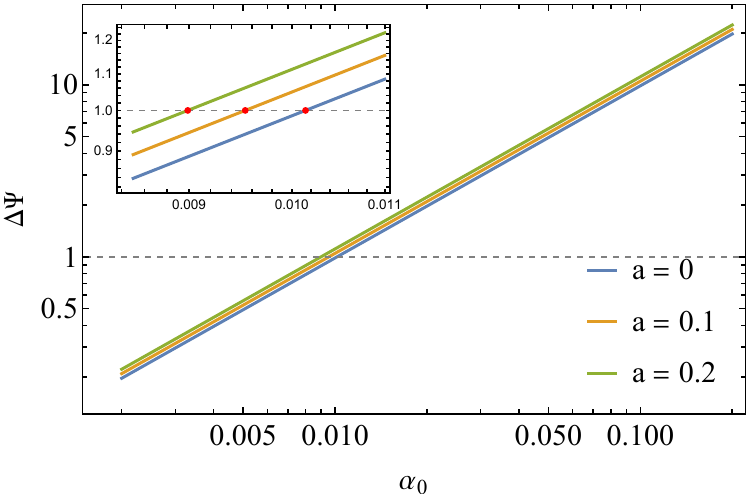}\hspace{0.2mm}	
	\caption{The orbital dephasing $\Delta \Psi$ as a function of $\alpha_0$ for different values of $a$. The gray dashed line corresponds to the LISA detection threshold of $\Delta \Psi\sim 1$ rad. }
	\label{Dephasing}
\end{figure}

Following the approach of Refs.~\cite{Cutler:1994pb, Glampedakis:2002ya}, we work within the adiabatic approximation, assuming that the particle moves along a geodesic and that the timescale of the orbital period is much shorter than the radiation-reaction timescale. In this framework, the averaged gravitational-wave energy and angular momentum fluxes are given by
\begin{eqnarray}
\label{GW_fluxes}
    \left\langle \frac{d E}{dt} \right \rangle=-\dot{E},\qquad \left\langle \frac{d L_z}{dt} \right\rangle=-\dot{L}_z.
\end{eqnarray}
Given that $E$ and $L$ are functions of $p$ and $e$, the evolution of the orbital parameters is governed by
\begin{eqnarray}\label{balance2}
\begin{split}
   -\dot{E}&=\frac{\partial E}{\partial p}\frac{dp}{dt}+\mu\frac{\partial E}{\partial e}\frac{de}{dt}, \\ 
-\dot{L}_z&=\frac{\partial L^z}{\partial p}\frac{dp}{dt}+\mu\frac{\partial L^z}{\partial e}\frac{de}{dt}.
\end{split}
\end{eqnarray}
Solving this system yields the averaged rates of change for $p$ and $e$:
\begin{equation}
\begin{split}
\frac{dp}{dt}&=\left(-\frac{\partial L_z}{\partial e}\dot{E}+\frac{\partial E}{\partial e}\dot{L}_z\right)\bigg/\left(\frac{\partial L_z}{\partial e}\frac{\partial E}{\partial p}-\frac{\partial E}{\partial e}\frac{\partial L_z}{\partial p}\right),\\
\frac{de}{dt}&=\left(\frac{\partial L_z}{\partial p}\dot{E}-\frac{\partial E}{\partial p}\dot{L}_z\right)\bigg/\left(\frac{\partial L_z}{\partial e}\frac{\partial E}{\partial p}-\frac{\partial E}{\partial e}\frac{\partial L_z}{\partial p}\right).
\end{split}
\end{equation}
To evaluate the observable implications of quantum corrections in the EMRI system, we compute the quadrupolar dephasing:
\begin{eqnarray} 
\label{quadrupole-dephasing}
\Delta\Psi_i=2\int_0^{T_{obs}}\Delta\Omega_i dt, \ \ \ \Delta\Omega_i=\Omega_i-\Omega_i^{GR},
\end{eqnarray}
where $i$ denotes the radial ($r$) or azimuthal ($\phi$) component.
Since the contribution from $\Delta\Psi_r$ is negligible compared to that of $\Delta\Psi_{\phi}$, the GW dephasing is dominated by $\Delta\Phi \sim \Delta\Psi_{\phi}$. Following Ref. \cite{Gupta:2021cno}, we adopt $\Delta\Phi\sim1$ rad as the minimum detectable dephasing for the LISA detector with SNR of $\sim 30$ \cite{Bonga:2019ycj}. Fig. \ref{Dephasing} presents preliminary estimates of the dephasing as a function of the quantum-gravity parameter $\alpha_0$ for various values of the spin $a$. It is observed that a larger $\alpha_0$ induces more significant deviations from GR, as manifested by more pronounced dephasing. Moreover, the dephasing is further enhanced with increasing $a$ for a fixed value of $\alpha_0$. 

\begin{figure}[H]
	\centering
        \includegraphics[width=0.45\textwidth]{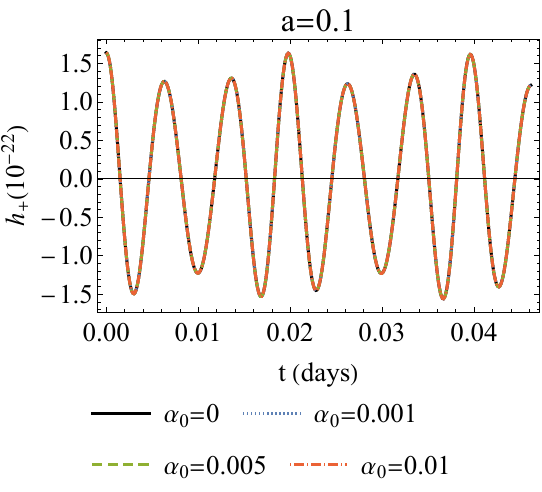}\hspace{0.2mm}	
 	\includegraphics[width=0.45\textwidth]{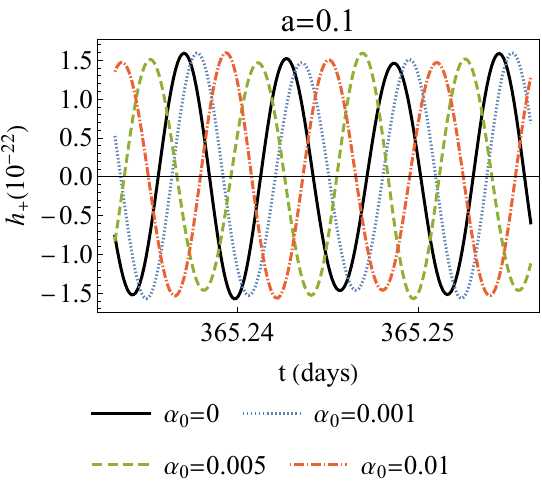}\hspace{0.2mm}	
	\caption{The comparison of the plus polarization $h_+$ in the AAK waveform across different values of $\alpha_0$. }
	\label{AAK_waveform}
\end{figure}

We now analyze the EMRI waveform. The augmented analytic kludge (AAK) model provides an efficient routine to generate the EMRI waveform and is widely used in LISA data analysis. For a detailed description of the AAK waveform model, see Refs. \cite{Chua:2017ujo, Chua:2015mua, PhysRevD, Fu:2024cfk, Zhang:2024csc}. Employing the \texttt{FastEMRIWaveforms (FEW)} package with fixed orbital parameters $\{a, p_0, e_0, M, \mu \}=\{0.1, 11, 0.1, 10^6 M_\odot, 10 M_\odot\}$, we generate the AAK waveform as presented in Fig. \ref{AAK_waveform}. The results demonstrate that after one-year observation, the quantum gravity effects induce significant deviations of the EMRI waveform from the corresponding Schwarzschild‑based waveform. This indicates that the EMRI system can serve as a potential probe for detecting signatures of quantum gravity in the space-base GW signals. 

\section{TIME-DELAY INTERFEROMETRY}\label{sec-4}

LISA, a space-based gravitational-wave detector, consists of three spacecrafts. Each pair of spacecrafts forms a Michelson-type interferometer via laser links. However, the motion of the spacecrafts makes it challenging to maintain equal arm lengths, which introduces the laser noise into the measurements. This issue is mitigated by time-delay interferometry (TDI) \cite{Cornish:2002rt,Cornish:2003tz,Tinto:1999yr,Tinto:2002de,Tinto:2014lxa,Vallisneri:2004bn,Marsat:2018oam,Armstrong:1999er, Estabrook:2000ef}, a technique that employs linear combinations of time-shifted measurements to suppress both laser noise and phase fluctuations.
Following the standard LISA conventions, the spacecrafts are labeled from $1$ to $3$ in clockwise order. Each Movable optical sub-assembly (MOSA) is denoted with two subscripts $ij$, where $i$ is the index of the local spacecraft and $j$ is that of the remote spacecraft. 

The first step involves computing the deformation induced in each of the six LISA links. This is achieved by projecting the GW strain along the unit vector $\hat{n}_{ij}$ of a given link. As an example, we consider the light traveling along link $12$ from the spacecraft 2 to spacecraft 1. The corresponding deformation $H_{12}(t)$ along this link is given by \cite{Katz:2022yqe,Baghi:2023qnq}
\begin{equation}
H_{12}(t) = h_+^\text{SSB}(t) \times \xi_+(\hat{u}, \hat{v}, \hat{n}_{12}) + h_\times^\text{SSB}(t) \times \xi_\times(\hat{u}, \hat{v}, \hat{n}_{12}),
\label{eq:projected-strain-ssb}
\end{equation}
where $\hat{u}$ and $\hat{v}$ are the polarization vectors of the GW source. The antenna pattern functions are
\begin{equation}
\begin{split}
    \xi_+(\hat{u}, \hat{v}, \hat{n}_{12}) &= (\hat{u} \cdot \hat{n}_{12})^2 - (\hat{v} \cdot \hat{n}_{12})^2, 
    \\
    \xi_\times(\hat{u}, \hat{v}, \hat{n}_{12}) &= 2 (\hat{u} \cdot \hat{n}_{12}) (\hat{v} \cdot \hat{n}_{12}).
\end{split}   
\end{equation}
When the light is emitted from the spacecraft 2 at $t_2$ and received at spacecraft 1 at time $t_1$, the reception time $t_1$ is related to $H_{12}(x,t)$ by
\begin{equation}
t_{1}\approx t_{2}+\frac{L_{12}}{c}-\frac{1}{2c}\int_{0}^{L_{12}}H_{12}({\hat{x}}(\lambda),t(\lambda))\,\mathrm{d}\lambda\,.
\end{equation}
Using the first-order approximations for the wave propagation time, $t(\lambda)\approx t_2+\lambda / c$, and wave position, $\hat{x}(\lambda)\approx\hat{x}_2(t_2)+\lambda\hat{n}_{12}(t_2)$, the deformation $H_{12}$ can be further refined as
\begin{eqnarray}
H_{12}(\hat{x}(\lambda),t(\lambda))&=&H_{12}\left(t(\lambda)-\frac{\hat{k}\cdot \hat{x}(\lambda)}{c}\right) \\  \nonumber
&=&H_{12}\left(t_2-\frac{\hat{k}\cdot \hat{x_2}(t_2)}{c}+\frac{1-\hat{k}\cdot\hat{n}_{12}(t_2)}{c}\lambda\right),
\end{eqnarray}
where $\hat{k}$ is the propagation vector in the  Solar system’s barycenter (SSB) frame. Defining the position of the receiver spacecraft at the reception time as $\hat{x}_1(t_1)=\hat{x}_2(t_2)+L_{12}\hat{n}_{12}$, and assuming that the spacecraft moves slowly compared to the GW propagation timescale, we find $\hat{x}_2(t_1) \approx \hat{x}_2(t_2)$ and $\hat{n}_{12}(t_1) \approx \hat{n}_{12}(t_2)$. The relative frequency shift, $y_{12}$, for the light traveling from spacecraft $2$ to spacecraft $1$ is given by
\begin{eqnarray}
y_{12}(t_1)\approx\frac{1}{2(1-\hat{k}\cdot\hat{n}_{12}(t_1))}\left[H_{12}\left(t_1-\frac{L_{12}(t_1)}{c}-\frac{\hat{k}\cdot\hat{x}_2(t_1)}{c}\right)-H_{12}\left(t_1-\frac{\hat{k}\cdot\hat{x}_1(t_1)}{c}\right)\right].  \nonumber \\ 
\end{eqnarray}
Here, $L_{12}$ represents the arm length between the spacecraft $2$ and spacecraft $1$. In this way, the first-generation Michelson TDI combinations, $X$, $Y$ and $Z$, are then constructed from the $y_{ij}$ time series along each of the six links 
\begin{equation*}
    \begin{split}
        X &=
        y_{13} + \delay{13} y_{31} + \delay{131} y_{12} + \delay{1312} y_{21} - [y_{12} + \delay{12} y_{21} + \delay{121} y_{13} + \delay{1213} y_{31}] ,\\ 
        Y &=
        y_{21} + \delay{21} y_{12} + \delay{212} y_{23} + \delay{2123} y_{32} - [y_{23} + \delay{23} y_{32} + \delay{232} y_{21} + \delay{2321} y_{12}] ,\\ 
        Z &=
        y_{32} + \delay{32} y_{23} + \delay{323} y_{31} + \delay{3231} y_{13} - [y_{31} + \delay{31} y_{13} + \delay{313} y_{32} + \delay{3132} y_{23}],
\end{split}
\end{equation*}
where the delay operator is defined as
\begin{equation}
    \delay{i_1, i_2, \dots, i_n} x(t) = x\left(t - \sum_{k=1}^{n-1}{L_{i_k i_{k+1}}/c}\right).
\end{equation}
The quasi-uncorrelated set of TDI variables $\{A, E, T\}$  is constructed from the $\{X, Y, Z\}$  linear combinations,  defined by \cite{Vallisneri:2004bn}
\begin{align}
    A =& \frac{1}{\sqrt{2}}\left(Z-X\right),  \\
    E =& \frac{1}{\sqrt{6}}\left(X-2Y+Z\right) ,\\
    T =&\frac{1}{\sqrt{3}}\left(X+Y+Z\right).
\end{align}
The noise-weighted inner product between two EMRI signals is actually represented as the sum over all channels
\begin{equation}
\left\langle a|b \right\rangle = \sum_{i = A,E,T}\left\langle a^i|b^i \right\rangle= \sum_{i = A,E,T}4 Re\int_{0}^{\infty}\frac{\Tilde{a}^{i}(f)^*\Tilde{b}^{i}(f)}{S_{n}^{i}(f)}df,
\end{equation}
where $S_{n}^{i}(f)$ denotes the power spectral density (PSD) for the channel $i$. For the $A$, $E$, $T$ channels, these are given by \cite{Marsat:2020rtl}
\begin{equation}
    \begin{aligned}
        S_{A, E} &= 8\sin^2 (2 \pi f L)\left(
        {[2 + \cos (2 \pi f L)] S_{\rm oms} +
        [6 + 2\cos (4\pi f L) + 4\cos (2\pi f L)] S_{\rm acc}}\right), \\
        S_{T} &= 32 \sin^2 (2 \pi f L) \sin^2 (\pi f L)
        [S_{\rm oms} + 4 \sin^2 (\pi f L) S_{\rm acc}],
    \end{aligned}
\end{equation}
with the underlying noise components
\begin{equation}
    \begin{aligned}
        \sqrt{S_{\rm oms}} &= 15\times10^{-12} \frac{2\pi f}c \sqrt{1+\left(\frac{2\times10^{-3}}f\right)^4} , \\
        \sqrt{S_{\rm acc}}&= \frac{3\times10^{-15}}{2\pi f c} \sqrt{1+\left(\frac{0.4\times10^{-3}}f\right)^2}\sqrt{1+\left(\frac f{8\times10^{-3}}\right)^4}.
    \end{aligned}
\end{equation}

We now turn to the Fisher Information Matrix (FIM) method to assess the potential for detecting signatures of quantum gravity effects. In the large SNR limit, the Fisher information matrix $\Gamma_{ij}$ takes the form
\begin{equation}\label{gamma_ij}
\Gamma_{i j}=\left\langle\left.\frac{\partial h}{\partial \xi_{i}}\right| \frac{\partial h}{\partial \xi_{j}}\right\rangle_{\xi=\hat{\xi}}.
\end{equation}
The phase space $\xi$ in the time domain consists of 11 parameters
\begin{eqnarray}
\xi=(\ln{M}, \ln{m}, a, p_0, e_0, \alpha_0, \theta_s, \phi_s, \theta_l, \phi_l, D_L).
\end{eqnarray}
The statistical errors on $\xi$ are obtained by inverting the Fisher information matrix given in Eq. \eqref{gamma_ij} 
\begin{eqnarray}
\sigma_i=\Sigma_{ii}^{1/2} \ \ \text {with}\ \ \Sigma_{ij}\equiv \left\langle\delta\xi_i \delta\xi_j\right\rangle=(\Gamma^{-1})_{ij}.
\end{eqnarray}
Because the $T$ channel responds much more weakly to the signal than the $A$ and $E$ channels \cite{Vallisneri:2004bn}, we restrict our analysis to the latter two. The total SNR and the  covariance matrix are therefore summed over the $A$ and $E$ channels:
\begin{eqnarray}
\rho&=&\sqrt{\rho_A^2+\rho_E^2}=\sqrt{\left\langle h_A\mid h_A\right\rangle+\left\langle h_E\mid h_E\right\rangle}, \\
\sigma_{\xi_i}^2&=&(\Gamma_A+\Gamma_E)_{ii}^{-1}.
\end{eqnarray}

\begin{figure}
    \centering
    \includegraphics[width=0.9\columnwidth]{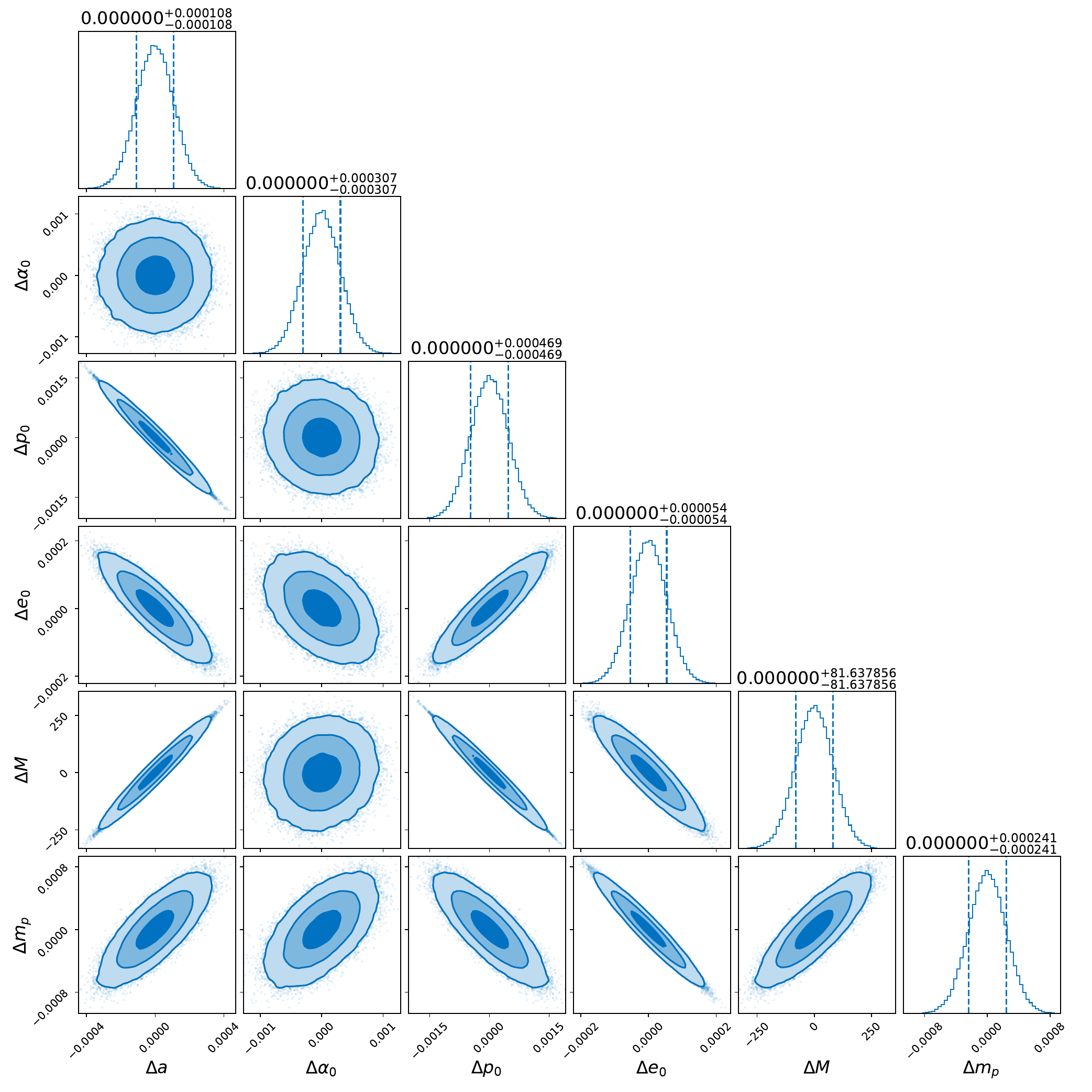}
    \caption{The posterior probability distributions for the intrinsic parameters for the channels of $A,E$. The plots along the diagonal show the marginalized distribution for each individual parameter, with vertical lines marking the $1\sigma$ credible interval. The two-dimensional contours represent the $68\%$, $95\%$, and $99\%$ joint credible intervals for parameter pairs.}
    \label{cornerTDI}
\end{figure}

To guarantee one year of adiabatic evolution prior to the final plunge, we selected the initial orbital separation $p_0$ and eccentricity $e_0$ accordingly, while adjusting the luminosity distance $D_L$ to achieve a signal-to-noise ratio of $150$.
The Fisher information matrix (FIM) analysis was performed on an EMRI system with $\mu=10~M_{\odot}$, $M=10^6~M_{\odot}$, $a=0.1$, $\alpha_0=0$, $p_0=9.24419$, $e_0=0.1$, and fixed source angles $\theta_s=\pi/3,~\phi_s=\pi/2$, $\theta_l=\pi/4,~\phi_l=\pi/4$.
Fig. \ref{cornerTDI} shows the probability distributions for the intrinsic parameters $\{a, \alpha_0, p_0, e_0, \ln M, \ln m_p\}$ of the regular black hole \eqref{metric}. This result suggests that the LISA would be capable of measuring departures from GR with an error of $\Delta\alpha_0=3.07\times10^{-4}$.

\section{Conclusions and Discussions}\label{sec-5}

In this work, we investigate an EMRI system consisting of a stellar-mass object orbiting a central spinning Hayward black hole. Due to quantum gravity effects, additional correction terms parameterized by $\alpha_0$ are incorporated, which modify the GR terms for the orbital frequency and gravitational radiation flux. Although the corrections from the quantum gravity only occur at higher post-Newtonian orders, their cumulative effect over the long inspiral timescale manifests as a detectable dephasing in the orbital waveform. Employing the \texttt{FastEMRIWaveforms} package, we generate AAK waveforms for different values of $\alpha_0$. The results show distinct deviations from the GR baseline for non-zero $\alpha_0$ demonstrating that EMRIs can serve as a natural laboratory for probing quantum gravity effects.

For space-based GW constellations like LISA, TianQin, and Taiji, the orbital motion of the spacecraft results in the inability to cancel laser noise. To this end, the technique of TDI must be employed. Following the process of Refs. \cite{Katz:2022yqe, Baghi:2023qnq}, we first compute the GW strain for all six interferometric links by projecting the GW polarizations onto the unit vectors along the LISA arms. These projections are then combined with time shifts to construct the TDI observables. Subsequently, we employ the FIM method, estimating the measurement uncertainty for $\alpha_0$ to be $\Delta\alpha_0=3.07\times10^{-4}$ with SNR$=150$. This result opens a promising window for using the space-based GW astronomy to perform high-precision tests of fundamental physics.

Utilizing EMRIs as probes of the fundamental properties of spacetime remains a highly promising research direction. In the present work, our analysis relies on the adiabatic approximation at the 0th post-adiabatic (0PA) order which neglects interactions between the point particle and the gravitational field. As noted in related studies \cite{Burke:2023lno}, this omission may introduce systematic biases in parameter estimation. Addressing these higher-order effects remains a significant theoretical challenge and will be the focus of our future work.

\acknowledgments

This work is supported by National Key R$\&$D Program of China (No. 2023YFC2206703), the Natural Science Foundation of China under Grant Nos. 12275079, 12505078, 12505085, 12375055, the China Postdoctoral Science Foundation (Grant No. 2025T180931), and the Jiangsu Funding Program for Excellent Postdoctoral Talent (Grant No. 2025ZB705).


\bibliographystyle{style1}
\bibliography{Ref}
\end{document}